\documentclass{ifacconf}
\usepackage{graphicx}
\usepackage{natbib}
\usepackage{amsmath,amssymb,amsfonts}
\usepackage{mathrsfs}
\usepackage{xspace}
\usepackage{color}
\usepackage{arydshln}
\usepackage{mathtools}

\usepackage{tikz}
\usepackage{xspace}

\DeclareRobustCommand{\legendline}[1]{\hspace{-3pt}
\tikz[#1,line width=0.4mm,baseline=-0.5ex]{\draw (0,0) -- (.35,0);}
\hspace{-3pt}}

\definecolor{mblue}{rgb}{0,0.4470,0.7410}
\definecolor{morange}{rgb}{0.8500,0.3250,0.0980}
\definecolor{myellow}{rgb}{0.9290,0.6940,0.1250}
\definecolor{mpurple}{rgb}{0.4940,0.1840,0.5560}
\definecolor{mgreen}{rgb}{0.4660,0.6740,0.1880}
\definecolor{mcyan}{rgb}{0.3010,0.7450,0.9330}
\definecolor{mred}{rgb}{0.6350,0.0780,0.1840}
\definecolor{mgreenblue}{rgb}{0.0,1.0,0.5}
\definecolor{parulablue}{rgb}{0.2431,0.1490,0.6588}
\definecolor{parulalblue}{RGB}{39,151,235}
\definecolor{parulagreen}{RGB}{129,204,89}
\definecolor{parulayellow}{RGB}{249,251,21}

\newcommand{\norm}[1]{\left\lVert#1\right\rVert}

\newcommand{\ltwo}{\ensuremath{\mathcal{L}_2}\xspace}

\newcommand{\litwo}{\ensuremath{\mathcal{L}_{\mathrm{i}2}}\xspace}
\newcommand{\m}[1]{\mathcal{#1}}
\newcommand{\mr}[1]{\mathrm{#1}}

\usepackage[left=43pt,right=43pt,top=106.75pt,bottom=39pt]{geometry}
\begin{document}
\begin{frontmatter}

\title{Linear Parameter-Varying Control of Nonlinear Systems based on \\Incremental Stability\thanksref{footnoteinfo}} 

\thanks[footnoteinfo]{This work has received funding from the European Research Council (ERC) under the European Unions Horizon 2020 research and innovation programme (grant agreement No 714663). \linebreak The results provided in this paper are the first steps towards incremental stability based LPV synthesis and further extensions and analysis is provided in \cite{Koelewijn2019NLTR}.}

\author[First]{P.J.W. Koelewijn} 
\author[First]{R. T\'oth}
\author{H. Nijmeijer\,$^{**}$}

\address[First]{Control Systems Group, Department of Electrical Engineering,  \textnormal{$^{**}$}\normalsize \,Dynamics and Control Group, Department of Mechanical Engineering, Eindhoven University of Technology, 5612 AE, Eindhoven, The Netherlands, (e-mail: \{p.j.w.koelewijn,\,r.toth,\,h.nijmeijer\}@tue.nl).}

\begin{abstract}The Linear Parameter-Varying (LPV) framework has long been used to guarantee performance and stability requirements of nonlinear (NL) systems mainly through the \ltwo-gain concept. However, recent research has pointed out that current \ltwo-gain based LPV synthesis methods can fail to guarantee these requirements if stabilization of a non-zero operating condition (e.g. reference tracking, constant disturbance rejection, etc.) is required. In this paper, an LPV based synthesis method is proposed which is able to guarantee incremental performance and stability of an NL system even with reference and disturbance rejection objectives. The developed approach and the current \ltwo LPV synthesis method are compared in a simulation study of the position control problem of a Duffing oscillator, showing performance improvements of the proposed method compared to the current \ltwo-based approach for tracking and disturbance rejection.
\end{abstract}

\end{frontmatter}

\section{Introduction}\label{sec:Introduction}
The Linear Parameter-Varying (LPV) framework has been developed in order to guarantee stability and performance requirements for nonlinear (NL) systems by extending the well-known synthesis results on guaranteeing these requirements for Linear Time-Invariant (LTI) systems, such as $H_\infty$-control, see e.g. \citep{Packard1994,Apkarian1995,Wu1995,Scherer2001} for some of the LPV synthesis results. It was thought that these results naturally extended the guarantees on tracking and disturbance rejection for NL systems through the LPV embedding concept. However, as exemplified in \citep{Scorletti2015}, this is not completely true. In \citep{Fromion1999a,Fromion2001,Scorletti2015} it is pointed out that a different notion of stability is required in order to also guarantee tracking and rejection requirements for NL systems, namely, the notion of incremental \ltwo-gain stability as opposed to \ltwo-gain stability that is currently used for LPV controller synthesis.

 Whereas the notion of $\ltwo$-gain stability only guarantees stability with respect to the origin of the system, incremental stability guarantees stability with respect to other trajectories. This notion of stability is therefore especially relevant for tracking and disturbance rejection, where the system has a non-zero operating condition. Similar notions to that of incremental stability were also developed, such as convergence, see \citep{Pavlov2004}, and contraction, see \citep{Lohmiller-And1998}. In the NL literature, various controller design methods have been developed to guarantee  convergence and contraction, see e.g. \citep{Pavlov2007,Lohmiller2000}. However, the resulting synthesis methods rely on complex procedures requiring expert knowledge in order to construct a stabilizing controller and allow no possibility for performance shaping, compared to LPV synthesis methods that offer powerful shaping paradigms.

A first attempt to guarantee incremental performance and stability of an NL system through the LPV framework was proposed in \citep{Scorletti2015}. A short coming of this method is that it can only be used for a subclass of NL control problems, namely ``filter cancelation control problems''. Moreover, the procedure requires a specific synthesis method to be used, resulting in a linear controller where scheduling-variable is only in an additive relationship with the controller input, limiting the obtainable performance.
Therefore, as main contribution of this paper, a synthesis method is proposed that is able to guarantee incremental stability and performance through the LPV framework for NL systems. The proposed method can be used for a larger class of NL control problems and systems. Furthermore, it allows the flexibility of using existing LPV synthesis methods during the synthesis procedure.

The paper is structured as follows. In Section \ref{sec:Problem}, a formal problem definition is given. Section \ref{sec:IncrFramework} describes the proposed solution. In Section \ref{sec:Examples}, the proposed synthesis method is applied to the position control problem of an NL Duffing oscillator. Finally, in Section \ref{sec:Conclusion}, conclusions are drawn on the developed results.

\subsubsection{Notation}
$\mathbb{R}$ is the set of real numbers, while $\mathbb{R}_+\subset\mathbb{R}$ is the set of non-negative reals. 
$\mathscr{L}_2^q$ is the space of square integrable real valued functions $\mathbb{R}_+ \rightarrow \mathbb{R}^q$, with the norm $\norm{f}_2 = \sqrt{\int_0^\infty \norm{f(t)}^2 dt}$, where $\norm{\star}$ is the Euclidean (vector) norm. 
A function is of class $\mathcal{C}_n$ if its first $n$ derivatives exist and are continuous. $\mathscr{L}\mathcal{C}^q_n$ denotes the set of all $\mathscr{L}_2^q$ functions that are in $\mathcal{C}_n$.
The identity matrix of size $N$ is denoted by $I_N$. The operator $\circ$ denotes the composition of two functions, i.e. $f\circ g = f(g)$.

\section{Problem Statement}\label{sec:Problem}
Consider a dynamical system 
given by
\begin{equation}\label{eq:nonlinsys}
\Sigma: \left \lbrace \begin{aligned}
\dot{x}(t) &= f\left(x(t),w(t)\right);\\
z(t) &= h\left(x(t),w(t)\right);\\
x(0) &= x_0;
\end{aligned} \right. 
\end{equation} 
where $x(t) \in \mathbb{R}^{n_\mathrm{x}}$ with $x_0 \in X \subseteq \mathbb{R}^{n_\mathrm{x}}$ is the state variable associated with the considered state-space representation of the system, $w \in \mathbb{R}^{n_\mathrm{w}}$ is the generalized disturbance, and $z \in \mathbb{R}^{n_\mathrm{z}}$ is the performance output of the system. $X$ is considered to be a compact set, while $f$ and $h$ are assumed to be bounded and sufficiently smooth maps $\mathbb{R}^{n_\mathrm{x}} \times \mathbb{R}^{n_\mathrm{w}}\rightarrow\mathbb{R}^{n_\mathrm{x}}$ and $\mathbb{R}^{n_\mathrm{x}} \times \mathbb{R}^{n_\mathrm{w}}\rightarrow\mathbb{R}^{n_\mathrm{z}}$ such that trajectories are unique and forward complete for all initial conditions $x_0 \in X$ and for all input functions $w \in \ltwo^{n_\mr{w}}$. Driven by the classical generalized plant concept, $w$ corresponds, as aforementioned, to the generalized disturbance channels (e.g. reference, external load, etc.) for which the performance of the systems is characterized by $z$ (e.g. tracking error, actuator usage, etc.).

As mentioned in the introduction, the current notion of stability used to guarantee stability and performance requirements of NL systems through the LPV framework is that of \ltwo-gain stability. For a dynamical system \eqref{eq:nonlinsys}, the notion of \ltwo-gain is given by the following definition.

\begin{defn}[\ltwo-gain]\label{def:l2gain}
$\Sigma$, given by \eqref{eq:nonlinsys}, is said to be \ltwo-gain stable if for all $w \in \mathscr{L}_{2}^{n_\mathrm{w}}$ and $x_0 \in X$, $\Sigma(w)$ exists and there is a finite $\gamma \geq 0$
and a bounded function $\zeta(x)\geq  0$ with $\zeta(0) = 0$  such that
\begin{equation}\label{eq:l2gain}
\norm{\Sigma(w)}_2 \leq \gamma \norm{w}_2+\zeta(x_0).
\end{equation}
The induced \ltwo-gain of $\Sigma$, denoted by $\norm{\Sigma}_2$, is the infimum of $\gamma$ such that \eqref{eq:l2gain} still holds.
\end{defn}
As exemplified in \citep{Scorletti2015}, this notion of stability is not able to guarantee tracking and rejection requirements for NL systems, as the $\ltwo$-gain only guarantees stability and performance with respect to the origin. When performing reference tracking and disturbance rejection, the system is in a non-zero operating condition, hence, $\ltwo$-gain stability is not the proper notion to use to also guarantee these requirements. Therefore, a different notion of stability has to be used, namely the notion of incremental stability.

The notion of incremental stability was first introduced in \citep{Zames1966} to provide conditions on continuity and stability of NL systems. For a dynamical system given by \eqref{eq:nonlinsys} the incremental-gain is given by the following defintion.

\begin{defn}[Incremental gain]\label{def:incrgain}
$\Sigma$, given by \eqref{eq:nonlinsys}, is said to be incrementally \ltwo-gain stable, from now on denoted as \litwo-gain stable, if it is \ltwo-gain stable and, there exist a finite $\eta \geq 0$ and a function $\zeta(x,\tilde{x})\geq  0$ with $\zeta(0,0) = 0$  such that
\begin{equation}\label{eq:incrml2gain}
\norm{\Sigma(w)-\Sigma(\tilde{w})}_2 \leq \eta \norm{w-\tilde{w}}_2+\zeta(x_0,\tilde{x}_0),
\end{equation}
for all $w,\tilde{w} \in \mathscr{L}_{2}^{n_\mathrm{w}}$ and $x_0,\tilde{x}_0 \in X$. The induced \litwo-gain of $\Sigma$, denoted by $\norm{\Sigma}_{\mathrm{i}2}$, is the infimum of  $\eta$ such that \eqref{eq:incrml2gain} holds, \citep{Fromion2003}. 
\end{defn} 
Based on this definition, \litwo-gain stability ensures convergence of system trajectories with respect to each other, whereas \ltwo-gain stability, as defined in Definition \ref{def:l2gain} can only ensure convergence with respect to one fixed point, e.g. the origin. Therefore, $\litwo$-gain stability can be used to also guarantee tracking and rejection requirements. In case an LTI system is considered, \ltwo-gain stability and \litwo-gain stability are equivalent (whereas this is not the case for NL systems), see \citep{Koelewijn2019IncrGain} for a proof. Consequently, the notion of $\ltwo$-gain \textit{can} be used to guarantee tracking and rejection requirements for LTI systems. Besides \litwo-gain stability, similar stability notions were also developed, namely contraction, see \citep{Lohmiller-And1998}, and convergence, see \citep{Pavlov2004}. 

One way to assess the incremental stability of a dynamical system given by \eqref{eq:nonlinsys} is by using the notion of the Gâteaux derviative of a system, first proposed in \citep{Fromion2003}.
\begin{defn}[G\^ateaux derivative]\label{def:gateaux}
Let $\Sigma$ defined by \eqref{eq:nonlinsys}, be such that $f$ and $h$ are $\mathcal{C}_1$. For a given input trajectory $w_\mathrm{r} \in \mathscr{L}_{2}^{n_\mathrm{w}}$ and initial condition $x_0\in X$, let $x_\mathrm{r}(t)$ be the solution of \eqref{eq:nonlinsys} for input $w_\mathrm{r}$ and $x_\mathrm{r}(0) = x_0$. The G\^ateaux derivative of $\Sigma$ is defined w.r.t. $(w_\mathrm{r},\,x_0)$ as 
\begin{equation}\label{eq:gateaux}
\delta \Sigma_{[w_\mathrm{r},x_0]}: \left \lbrace \begin{aligned}
\delta \dot{{x}}(t) &= A(t)\delta x(t) + B(t) \delta w(t);\\
\delta z(t) &= C(t)\delta x(t) + D(t) \delta w(t);\\
\delta x(0) &= \delta x_0; 
\end{aligned}\right.  
\end{equation}
with $A(t) = \frac{\partial f}{\partial x}(x_\mathrm{r}(t),w_\mathrm{r}(t))$, $B(t) = \frac{\partial f}{\partial w}(x_\mathrm{r}(t),w_\mathrm{r}(t))$, $C(t) = \frac{\partial h}{\partial x}(x_\mathrm{r}(t),w_\mathrm{r}(t))$, $D(t) = \frac{\partial h}{\partial w}(x_\mathrm{r}(t),w_\mathrm{r}(t))$ be bounded functions, $(\delta w,\delta x, \delta z)\in (\mathscr{L}_{2}^{n_\mathrm{w}} \times \mathscr{L}_2^{n_\mathrm{x}} \times \mathscr{L}_2^{n_\mathrm{z}})$ and $\delta x_0 \in \mathbb{R}^{n_\mathrm{x}}$, \citep{Fromion2003}.
\end{defn}
From now on, a system given by \eqref{eq:gateaux} will be referred to as the incremental form of the NL system \eqref{eq:nonlinsys}, while the original NL system will be referred to as the primal form.
Using this definition, the following theorem is given in \citep{Fromion2003} to assess the incremental stability of an NL system.
\begin{thm}\label{thrm:incrgateaux}
Let $\Sigma$, as defined in \eqref{eq:nonlinsys}, be G\^ateaux differentiable in terms of Definition \ref{def:gateaux}, meaning that $\delta \Sigma_{[w_\mathrm{r},x_0]}$ exists for any $w_\mathrm{r} \in \mathscr{L}_{2}^{n_\mathrm{w}}$ and $x_0\in X$. Then, $\norm{\Sigma}_{\mathrm{i}2} \leq \eta$, if and only if $\norm{\delta\Sigma_{[w_\mathrm{r},x_0]}}_{2} \leq \eta$ for any $w_\mathrm{r} \in \mathscr{L}_{2}^{n_\mathrm{w}}$ and $x_0\in X$.\end{thm} 
For a proof see \citep{Fromion2003}. In short, under our assumptions, the induced \ltwo-gain of the incremental form of system is equal to the induced \litwo-gain of the primal form of the system. In \citep{Scorletti2015}, Theorem \ref{thrm:incrgateaux} is used in conjunction with the LPV framework to make the first steps toward synthesizing controllers ensuring incremental stability of NL systems.

In this paper, it is proposed that using Theorem \ref{thrm:incrgateaux}, instead of directly synthesizing a controller guaranteeing \litwo-gain stability for an NL system, standard (\ltwo) LPV synthesis can be performed  on the (LPV embedded) incremental form of the system. This results in the incremental form of an LPV controller, which then needs to be transformed back to the its primal form in order to be used with the NL system. Doing this transformation is an issue also encountered in previous work. In \citep{Scorletti2015} this issue is circumvented by synthesizing a controller which has an additive dependency on the scheduling-variable, i.e. an LTI controller which has as an input the scheduling-variable. Because the incremental form and primal form of an LTI system are equivalent, no transformation is required to obtain the primal form of the controller.
In this paper we propose a new method in order to realize the primal form of the controller based on the synthesized incremental form of the LPV controller.


\section{Incremental LPV Controller}\label{sec:IncrFramework}
\subsection{Main Concept}
To formulate a control synthesis problem, the NL system is considered to have the form of a generalized plant:
\begin{equation}\label{eq:gensys}
P:\left\lbrace 
\begin{aligned}
	\dot{x} &= f(x(t),u(t)) + B_1 w(t)\\
	z &= h_1(x(t),u(t))+D_{11}w(t)\\
	y &= h_2(x(t))+D_{21}w(t)
\end{aligned}\right. 
\end{equation}
where $y \in \mathbb{R}^{n_y}$ is the measured output and $u \in \mathbb{R}^{n_u}$ the control input of the system, while $w$ and $z$ retain their roles in characterizing the performance channels. The functions $f$, $h_1$ and $h_2$ are considered to be $\m{C}_1$ such that the G\^ateaux derivative of \eqref{eq:gensys} exists. 

The main concept behind our proposed procedure is as follows:
\begin{enumerate}
    \item Compute the incremental form of the nonlinear system \eqref{eq:gensys};
    \item The incremental form of the system is then embedded in an LPV representation. For this LPV model, a controller is synthesized which ensures a minimal \ltwo-gain of the interconnection of the controller and plant from $\delta w$ to $\delta z$, which by Theorem \ref{thrm:incrgateaux} ensures \litwo-gain stability of interconnection of the primal form and (to be constructed) primal form of the controller from $w$ to $z$. This step can be accomplished by using standard synthesis procedures in the LPV framework, e.g. see \citep{Packard1994,Apkarian1995,Wu1995,Scherer2001}. This step will be referred to as the \emph{incremental synthesis} step;
    \item Finally, the synthesized controller of the previous step, which is in its incremental form, is realized back to its primal form for it to be used with the NL system. This step will be referred to as the \emph{controller realization} step. 
\end{enumerate}

\subsection{Incremental form computation}
\label{sec:genplantconstr}
As a first step in the procedure, the incremental form of the NL system \eqref{eq:gensys} is computed based on the G\^ateaux derivative. This results in a system of the form
\begin{equation}\label{eq:genplantincr}
\delta P_{[u_\mr{r},x_0]}: \left \lbrace
    \begin{aligned}
    \delta\dot{{x}}(t) &= \bar{A}({x}_{\mr{r}}(t),{u}_{\mr{r}}(t))\delta x(t) +{B}_1\delta w(t) \\ &\phantom{=}+\bar{B}_2({x}_{\mr{r}}(t),{u}_{\mr{r}}(t)) \delta {u}(t);\\
    \delta z(t) &= \bar{C}_1({x}_{\mr{r}}(t),{u}_{\mr{r}}(t))\delta {x}(t) +{D}_{11}\delta w(t) \\ &\phantom{=}+ \bar{D}_{12}({x}_{\mr{r}}(t),{u}_{\mr{r}}(t)) \delta {u}(t);\\
    \delta {y}(t) &= \bar{C}_2({x}_{\mr{r}}(t)) \delta {x}(t) +{D}_{21}\delta w(t);\\
    \delta {x}(0) &= \delta x_0;
\end{aligned}\right. 
\end{equation} 
where $\delta {x}$ is the incremental state, $\delta {u}$ is the incremental control input, $\delta {y}$ is the incremental measured output, $\delta w$ is the incremental generalized disturbance and $\delta z$ is the incremental performance output. Furthermore, $\bar{A}(x_{\mr{r}},u_{\mr{r}}) = \frac{\partial f}{\partial x}(x_{\mr{r}},u_{\mr{r}})$, $\bar{B}_2(x_{\mr{r}},u_{\mr{r}}) = \frac{\partial f}{\partial u}(x_{\mr{r}},u_{\mr{r}})$, $\bar{C}_1(x_{\mr{r}},u_{\mr{r}}) = \frac{\partial h_1}{\partial x}(x_{\mr{r}},u_{\mr{r}})$, $\bar{D}_{12}(x_{\mr{r}},u_{\mr{r}}) = \frac{\partial h_1}{\partial u}(x_{\mr{r}},u_{\mr{r}})$ and $\bar{C}_2(x_{\mr{r}},u_{\mr{r}}) = \frac{\partial h_2}{\partial x}(x_{\mr{r}},u_{\mr{r}})$. Moreover, it is assumed that this interconnection is well-posed.

\subsection{Incremental synthesis}\label{sec:incrsynth}
To synthesize an LPV controller, the incremental form of the plant $\delta P$ is embedded in an LPV representation. Embedding the incremental form \eqref{eq:genplantincr} in an LPV representation results in
\begin{equation}\label{eq:genplantincrLPV}
\delta P_\mathrm{LPV}: \left \lbrace
    \begin{alignedat}{3}
    \delta\dot{{x}}(t) &= A(\rho(t))\delta {x}(t) +B_1\delta w(t)\\ &\phantom{=}+B_2(\rho(t)) \delta {u}(t);\\
    \delta z(t) &= C_1(\rho(t))\delta {x}(t) +D_{11}\delta w(t) \\ &\phantom{=}+ D_{12}(\rho(t)) \delta {u}(t);\\
    \delta {y}(t) &= C_2(\rho(t)) \delta {x}(t) +D_{21}\delta w(t);\\
    \delta {x}(0) &= \delta x_0;
\end{alignedat}\right. 
\end{equation}
where $\rho(t) \in \mathcal{P} \subset \mathbb{R}^{n_{\rho}}$ is assumed to be measurable and $\m{P}$ is chosen to be a convex set. Moreover, there exists a function $\psi :\mathbb{R}^{n_\mr{x}} \times \mathbb{R}^{n_\mr{u}} \rightarrow \mathbb{R}^{n_{\rho}}$, such that $\rho(t) = \psi(x_\mr{r}(t),u_\mr{r}(t))$ and it is assumed that $x_r\in \m{X}\subset \mathbb{R}^{n_\mr{x}}$ and $u_r\in \m{U}\subset \mathbb{R}^{n_\mr{u}}$ such that $\psi(\m{X},\m{U})\subseteq \m{P}$. Therefore, we have the relations $\bar{A} = A \circ \psi$, $\bar{B}_2 = B_2 \circ \psi$, etc.
 Several methods for embedding exists, see e.g. \citep{Kwiatkowski2006AutomatedGA,toth2010modeling}. Embedding of \eqref{eq:genplantincr} is straightforward as it is already in a factorized form. 
Based on \eqref{eq:genplantincrLPV}, a controller is synthesized. For this part, \ltwo-gain based LPV synthesis procedures can be used, such as grid-based, polytopic, or LFT LPV synthesis techniques, e.g. \citep{Packard1994,Apkarian1995,Wu1995,Scherer2001}. 

Synthesizing a controller for \eqref{eq:genplantincrLPV} that ensures a certain \ltwo-gain stability and performance bound on the closed-loop interconnection of \eqref{eq:genplantincrLPV} and the synthesized controller results in an LPV controller $\delta K$. This LPV controller $\delta K$ is of the form
\begin{equation}\label{eq:incrContr}
\delta K: \left \lbrace
    \begin{alignedat}{2}
    \delta \dot{x}_\mathrm{k}(t) &= A_\mathrm{k}(\rho(t)) \delta x_\mathrm{k}(t) &&+ B_\mathrm{k}\delta u_\mathrm{k}(t);\\
    \delta y_\mathrm{k}(t) &= C_\mathrm{k}(\rho(t)) \delta x_\mathrm{k}(t) &&+ D_\mathrm{k} \delta u_\mathrm{k}(t);
    \end{alignedat}\right. 
\end{equation}
which will be referred to as the incremental form of the controller\footnote{Note that it is assumed that $B_\mathrm{k}$ and $D_\mathrm{k}$ are independent of the scheduling-variable, which needs to be ensured during synthesis. This property is exploited during the \textit{controller realization} step.}, where $\delta x_\mathrm{k}(t) \in \mathbb{R}^{n_\mathrm{x_k}}$, $\delta u_\mathrm{k}(t) \in \mathbb{R}^{n_{\mathrm{u_k}}}$ and $\delta y_\mathrm{k}(t) \in \mathbb{R}^{n_{\mathrm{y_k}}}$ are the states, inputs and outputs of the controller, respectively. The interconnection of controller and plant is such that $\delta u = \delta y_k$ and $\delta u_k = \delta y$.

Based on this synthesis step, it is then ensured that the closed-loop interconnection of the (incremental form of the) controller $\delta K$ and the incremental form of the plant $\delta P$ is \ltwo-gain stable for any $\rho(t) \in \m{P}$. Consequently, as per Theorem \ref{thrm:incrgateaux}, this guarantees the \litwo-gain stability of the primal form of the generalized plant $P$ interconnect with the to be constructed primal form of the controller (denoted by $K$) for all $\rho(t) \in \m{P}$, i.e. for all trajectories $(x,u)$ that remain in $(\m{X}\times \m{U})$. Note that due to the fact that incremental stability ensures contraction of the state trajectories of the closed-loop system, hence, theoretically $(x,u)$ is guaranteed to remain in $(\m{X}\times \m{U})$ under $w(t)\equiv 0$. However for $w(t)\neq 0$, it has to be verified posteriori that $(x,u)$ remain in $(\m{X}\times \m{U})$, i.e. all $\rho(t)$ remain in $\m{P}$, which is a limitation due to the use of the LPV framework.

\subsection{Controller realization}\label{sec:contrrealiz}
In order to realize the primal form of the controller $K$, based on its incremental form $\delta K$ in \eqref{eq:incrContr}, concepts related to velocity based linearizations are used.

\begin{figure}
\centering
\hspace{-20pt}
\includegraphics[scale=.9]{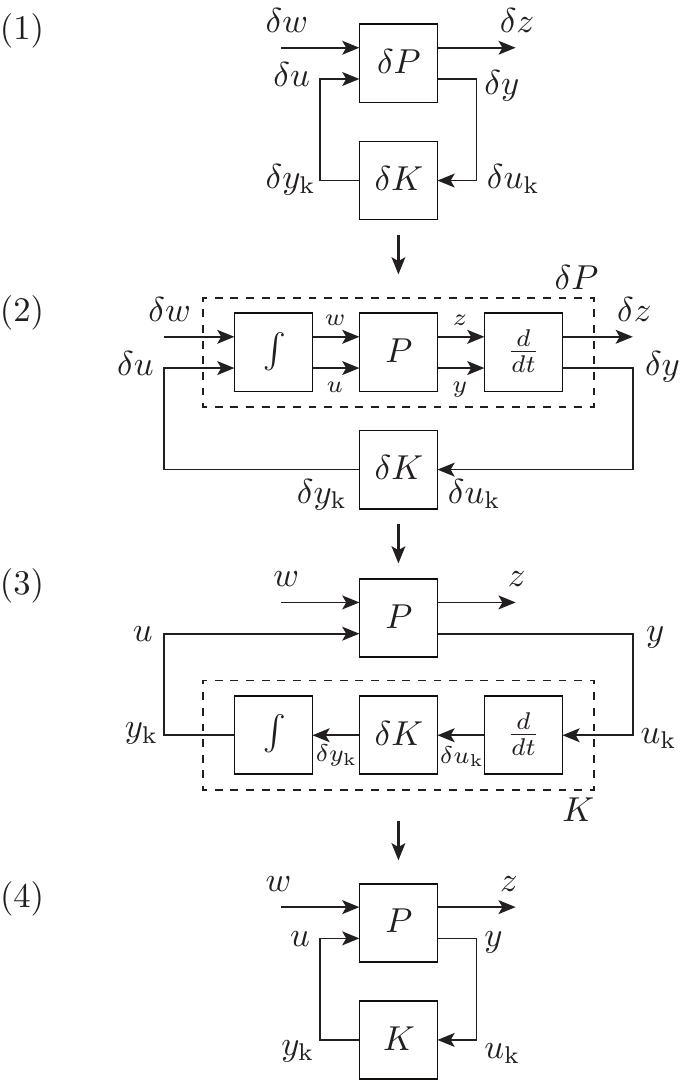}
\caption{Controller realization step.}
\label{fig:cntrRealiz}
\end{figure}

Under the restriction of considered solutions to $w \in \mathscr{L}\m{C}_1^{n_\mr{w}}$, $u \in \mathscr{L}\m{C}_1^{n_\mr{w}}$, $y \in \m{C}_1$ and $z\in\m{C}_1$ the velocity based linearization of a system, see \citep{Leith1999}, is equivalent with the G\^ateaux derivative of a system \eqref{eq:gateaux}
, the incremental form of a system can then be expressed by an interconnection of the primal form and integrators and differentiators. This relation can then be exploited to construct the controller as is shown by means of the following steps (also displayed in Fig. \ref{fig:cntrRealiz}):
\begin{enumerate}
    \item Assume that an incremental form of the controller $\delta K$ is synthesized for incremental form of the plant $\delta P$ using the procedures in Sections \ref{sec:genplantconstr} and \ref{sec:incrsynth};
    \item The incremental form of the plant $\delta P$ is then realized by, integrating the input of the primal form $P$ and differentiating the output;
    \item The integrator and differentiator are then moved to the side of the controller $\delta K$. 
    \item As the action of differentiation is problematic for noisy measurements, the overall controller is realized based on LPV realization theory. This results in the primal form $K$ of the controller where the integration and differentiation action are embedded in the controller.
\end{enumerate}

To realize the primal form of the controller for the fourth step, the integrator and differentiator together with the incremental form of the controller $\delta K$ are realized in one structure. To achieve this, the inputs of the controller need to be differentiated and the outputs need to be integrated, this results in the following relation
 \begin{equation}
 \begin{aligned}
 	\xi y_\mathrm{k} &= \delta y_\mathrm{k},\\
 	\xi u_\mathrm{k} &= \delta u_\mathrm{k},
 \end{aligned}
 \end{equation} 
 where $\xi$ denotes the operator $\xi = \frac{\partial}{\partial t}$ (note that this operator is non-commutative).
Using this relation together with \eqref{eq:incrContr} results in
\begin{equation}\label{eq:deltaK}\left\{
\begin{aligned}
    \xi \delta x_\mathrm{k} &= A_\mathrm{k}(\rho)\delta x_\mathrm{k}+B_\mathrm{k} \xi u_\mathrm{k};\\
    \xi y_\mathrm{k} &= C_\mathrm{k}(\rho)\delta x_\mathrm{k}+D_\mathrm{k} \xi u_\mathrm{k};
\end{aligned}\right.
\end{equation}
We will first rewrite the first equation of \eqref{eq:deltaK}, by using that $(\xi B_\mathrm{k})u_\mathrm{k}=0$, as
\begin{equation}
\begin{alignedat}{2}
    \xi \delta x_\mathrm{k} &= &&A_\mathrm{k}(\rho)\delta x_\mathrm{k}+ B_\mathrm{k}\xi u_\mathrm{k} + \left(\xi B_\mathrm{k}\right)u_\mathrm{k} ,\\
    \xi \left(\delta x_\mathrm{k}-B_\mathrm{k}u_\mathrm{k}\right) &= &&A_\mathrm{k}(\rho)\delta x_\mathrm{k} ,
\end{alignedat}
\end{equation}
we then define $\tilde{x}_\mathrm{k} = \delta x_\mathrm{k} - B_\mathrm{k}u_\mathrm{k}$, resulting in
\begin{equation}
\begin{aligned}
    \xi \tilde{x}_\mathrm{k} &= A_\mathrm{k}(\rho)\tilde{x}_\mathrm{k}+A_\mathrm{k}(\rho)B_\mathrm{k}u_\mathrm{k},\\
    \dot{\tilde{x}}_\mathrm{k} &= A_\mathrm{k}(\rho)\tilde{x}_\mathrm{k}+A_\mathrm{k}(\rho)B_\mathrm{k}u_\mathrm{k}.
\end{aligned}
\end{equation}
A similar procedure can be applied to the second equation of \eqref{eq:deltaK} in order to obtain
\begin{equation}
    \dot{\hat{x}}_\mathrm{k} = C_\mathrm{k}(\rho)\tilde{x}_\mathrm{k}+C_\mathrm{k}(\rho)B_\mathrm{k}u_\mathrm{k},
\end{equation}
where $\hat{x}_\mathrm{k} = y_\mathrm{k}-D_\mathrm{k}u_\mathrm{k}$. These two results can then be combined to obtain the complete controller $K$, resulting in
\begin{equation}\label{eq:controlller}
    \left[\begin{array}{c}
    \dot{\tilde{x}}_\mathrm{k}\\\dot{\hat{x}}_\mathrm{k} \\ \hdashline[2pt/2pt] y_\mathrm{k}
    \end{array}\right]=
    \left[\begin{array}{cc;{2pt/2pt}c}
    A_\mathrm{k}(\rho) & 0 & A_\mathrm{k}(\rho)B_\mathrm{k}\\C_\mathrm{k}(\rho) & 0 &C_\mathrm{k}(\rho)B_\mathrm{k}\\\hdashline[2pt/2pt]
    0 & I & D_k
    \end{array}\right]
    \left[\begin{array}{c}
    \tilde{x}_\mathrm{k}\\\hat{x}_\mathrm{k}\\\hdashline[2pt/2pt] u_\mathrm{k}
    \end{array}\right],
\end{equation}
which will be referred to as an \litwo (gain optimal) LPV controller or an incremental LPV controller\footnote{Due to the realization concept one can argue that stability and performance guarantees only apply to the primal closed-loop system if the signal trajectories are guaranteed to be in $\m{C}_1$. However, as the G\^ateaux derivative of the closed-loop is the interconnection of \eqref{eq:genplantincr} and \eqref{eq:incrContr}, hence, in terms of Theorem \ref{thrm:incrgateaux}, \litwo stability and performance holds without any restrictions.}. Note that  while the incremental LPV controller naturally has integral action, this is not the (only) cause of the improved tracking and rejection performance. In \cite{Scorletti2015} it is demonstrated that even an (\ltwo based) LPV controller with explicit integral action can fail to guarantee tracking and disturbance rejection requirements.

\section{Example}\label{sec:Examples}
In this section, the \litwo LPV controller synthesis method as described in the last section will be demonstrated with the position control problem of a Duffing oscillator. First, a standard \ltwo-gain LPV controller design will be synthesized, for which it will be shown that it can fail to adhere to the tracking and rejection requirements when a constant input disturbance is applied.
\subsection{Duffing Oscillator Dynamics}
A Duffing oscillator is a mass-spring-damper system which has a spring that generates a restoring force which is a cubic function of its displacement. The dynamics of a Duffing oscillator can be represented by the following NL state-space equations
\begin{equation}\label{eq:duffNL}
    \left\lbrace
    \begin{aligned}
        \dot{x}_1(t) &= x_2(t);\\
        \dot{x}_2(t) &= -\frac{k_1}{m} x_1(t) -\frac{k_2}{m} \left(x_1(t)\right)^3 - \frac{d}{m} x_2(t) + \frac{1}{m} u(t);\\
        y(t) &= x_1(t);
    \end{aligned}
    \right.
\end{equation}
where the (physical) parameters are the mass $m = 1 \,[\mathrm{kg}]$, the linear spring constant $k_1 = 0.5 \,[\mathrm{N}\cdot \mathrm{m}^{-1}]$, nonlinear spring constant $k_2 = 5\,[\mathrm{N}\cdot \mathrm{m}^{-3}]$ and damping coefficient $d = 0.2\,[\mathrm{N}\cdot \mathrm{s}\cdot \mathrm{m}^{-1}]$. 

\subsection{\ltwo-gain LPV Synthesis}\label{sec:l2gainsynth}
For \ltwo-gain LPV synthesis, \eqref{eq:duffNL} is embedded in an LPV representation, resulting in the following LPV model
\begin{equation}\label{eq:duffLPV}
    \left\lbrace
    \begin{aligned}
        \dot{x}_1 &= x_2;\\
        \dot{x}_2 &= \left(-\frac{k_1}{m}-\frac{k_2}{m}{\rho}\right) x_1 - \frac{d}{m} x_2 + \frac{1}{m} u;\\
        y &= x_1;
    \end{aligned}
    \right.
\end{equation}
where ${\rho} = y^2$ is the scheduling-variable, with $\rho \in \mathcal{P}=[0,\, 2]$, allowing a relatively large operating range. 

\begin{figure}
    \centering
    \includegraphics[scale=.9]{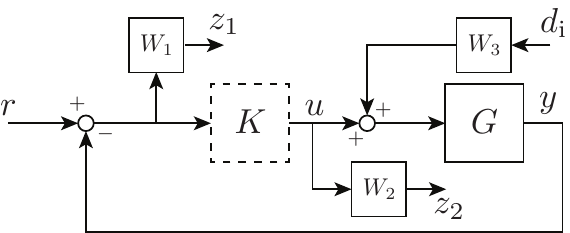}
    \caption{Generalized plant.}
    \label{fig:genplantscor}
\end{figure}

Based on the LPV model representation of the system, a generalized plant is constructed in order to achieve output reference tracking and input disturbance rejection. The generalized plant is shown in Fig. \ref{fig:genplantscor}, where $r$ and $d_\mathrm{i}$ are the reference and input disturbance respectively, together forming the disturbance channel $w$; $z_1$ and $z_2$ denote the performance channels; and $u$ and $y$ denote the control input and measured output of the generalized plant respectively. The weighting filter $W_1$ is designed with low-pass characteristics in order to have sufficient tracking performance at low frequencies, and has as transfer function $W_1(s) = \frac{0.5012 s + 2.506}{s + 2.506\cdot 10^{-4}}$. The weighting filter $W_2$ is designed with high-pass characteristics in order to have roll-off at high frequencies for the control input, and has as transfer function $W_2(s) = \frac{10 s + 800}{s + 8\cdot 10^{4}}$. Finally, the weighting filter $W_3$ is chosen as a constant gain, given by $W_3 = 1.5$.

Using this generalized plant, an \ltwo-gain optimal LPV controller is synthesized using the polytopic synthesis method based on \citep{Apkarian1998} where for the quadratic stability and performance condition, $X$ is considered parameter-varying and $Y$ to be constant. Synthesizing the LPV controller using this method results in an \ltwo-gain of 0.91.

\subsection{\litwo-gain LPV Synthesis}
In order to perform the incremental synthesis method as described in Section \ref{sec:IncrFramework}, the incremental form of \eqref{eq:duffNL} is computed and the resulting relation is embedded in an LPV representation, resulting in
\begin{equation}\label{eq:duffincr}
    \left\lbrace
    \begin{aligned}
        \delta \dot{x}_1 &= \delta x_2;\\
        \delta \dot{x}_2 &= \left(-\frac{k_1}{m}-3 \frac{k_2}{m} \rho\right) \delta x_1 - \frac{d}{m} \delta x_2 + \frac{1}{m} \delta u;\\
        \delta y &= \delta x_1;
    \end{aligned}
    \right.
\end{equation}
where again $\rho = y^2$ with bounds same as for \eqref{eq:duffLPV}. Note that the primal and incremental form can be embedded using the same scheduling map.

The same generalized plant structure is used as in Section \ref{sec:l2gainsynth}, see Fig. \ref{fig:genplantscor}, as well as the same LPV synthesis method, i.e. the polytopic synthesis method based on \citep{Apkarian1998}.
Performing the synthesis results in an \litwo-gain of 0.98. The resulting controller is realized using the procedure as described in Section \ref{sec:contrrealiz}.

\subsection{Simulation Results}

\begin{figure}
    \centering
    \includegraphics[scale=1]{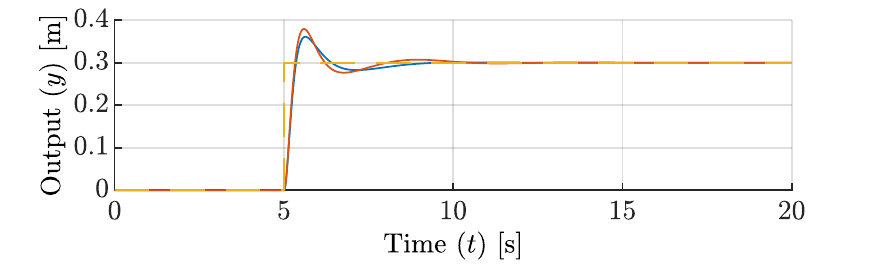}
    \caption{Tracking performance of the \ltwo controller (\legendline{mblue}), and \litwo controlller (\legendline{morange}); reference trajectory (\legendline{myellow,dashed}).} 
    \label{fig:duff_nodist}
\end{figure}

\begin{figure}
    \centering
    \includegraphics[scale=1]{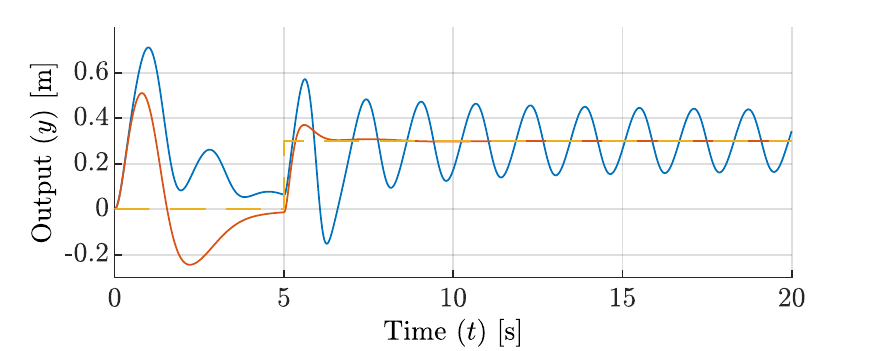}
    \caption{Tracking performance (with constant input disturbance) of the \ltwo controller (\legendline{mblue}), and \litwo controlller\protect\linebreak (\legendline{morange}); reference trajectory (\legendline{myellow,dashed}).} 
    \label{fig:duff_dist}
\end{figure}

\begin{figure}
    \centering
    \includegraphics[scale=1]{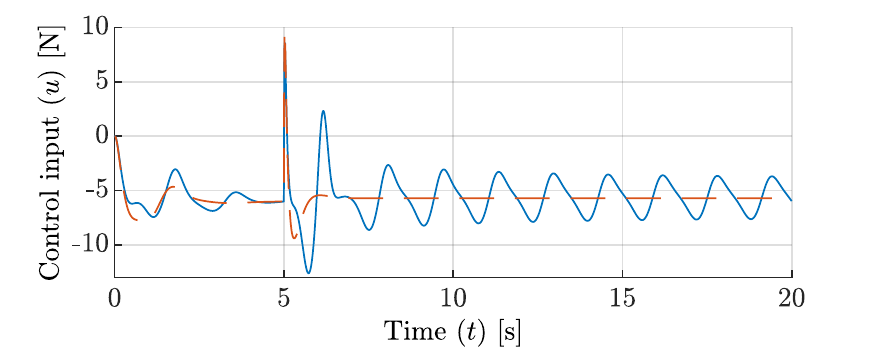}
    \caption{Generated control input  of the  \ltwo controller \protect\linebreak (\legendline{mblue}), and \litwo controlller (\legendline{morange,dashed}) (with constant input disturbance present).} 
    \label{fig:duff_input}
\end{figure}

Using the synthesized \ltwo LPV controller and \litwo LPV controller, the NL system \eqref{eq:duffNL} is simulated interconnected to the controllers. As a reference trajectory, a step signal is chosen which changes from 0 to 0.3 at $t=5$s. Furthermore, the system is also simulated with and without a constant input disturbance of 6 N. This input disturbance can for instance be seen as additional mass attached to the system. In Fig. \ref{fig:duff_nodist} the trajectories of the system interconnect with the respective controller are shown in the case when no input disturbance is present and in Fig. \ref{fig:duff_dist} the case when an input disturbance \textit{is} present is shown. Moreover, Fig. \ref{fig:duff_input} displays the control input generated by the controllers in the case where input disturbance is present.

From Fig. \ref{fig:duff_nodist} it can be observed that both the \ltwo and \litwo LPV controllers obtain very similar performance; the \litwo LPV controller only has slightly more overshoot, but also has a slightly lower settling time. On the other hand, when the input disturbance is present, shown in Fig. \ref{fig:duff_dist}, it is apparent that the \ltwo LPV controller is not able to guarantee the tracking and disturbance requirements anymore and depicts oscillatory behavior, similar to results described in \citep{Scorletti2015}. The proposed \litwo LPV controller design on the other hand is still able to obtain the desired performance. From Fig. \ref{fig:duff_input} it can clearly be seen that the oscillations are caused by the control input generated by the \ltwo LPV controller, as oscillations are also present in the control input signal, whereas this is not the case with the $\litwo$ LPV controller. 


\subsubsection{Analysis}
\begin{figure}
    \centering
    \includegraphics[scale=1]{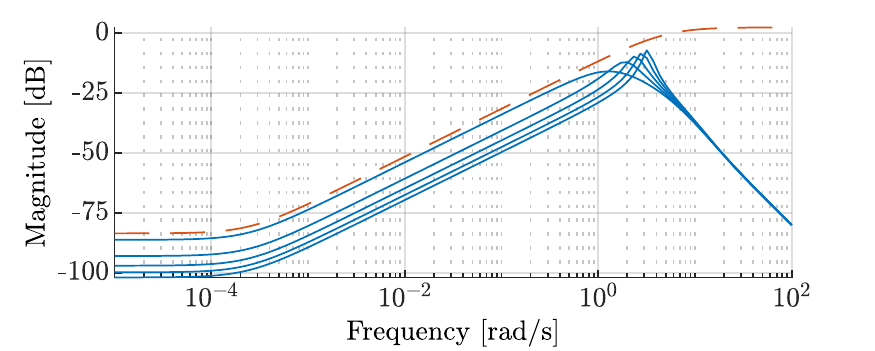} 
    \caption{Process sensitivity with $\ltwo$ LPV controller (\legendline{mblue}), and the corresponding inverse weighting filter (\legendline{morange,dashed}).} 
    \label{fig:bode_d_to_e}
\end{figure}

To show that the behavior of the \ltwo LPV controller is not due to improper tuning of the weighting filters, the process sensitivity (i.e. $d_\mathrm{i}$ to $z_1$) of the LPV plant \eqref{eq:duffLPV} in closed-loop interconnection with the \ltwo LPV controller is displayed in Fig. \ref{fig:bode_d_to_e} for frozen values of the scheduling-variable, along with the inverse weighting filter $W_1^{-1}\cdot W_3^{-1}$. Based on the graph in Fig. \ref{fig:bode_d_to_e} it would be natural to expect that if a constant input disturbance would be applied, the error is around -80 dB (this will be slightly higher due to also performing reference tracking simultaneously). In the generalized plant $W_3$ is chosen with a weight of 1.5, therefore when an input disturbance of 6 N is applied to the system, the expected error would be 4 times higher. Hence, based on the standard analysis of the \ltwo LPV controller, a constant tracking error of only -68 dB would be expected. This is in large contrast to the tracking error of the \ltwo LPV controller, seen in Fig. \ref{fig:duff_dist}, which has oscillatory behavior. This again highlights that the current analysis for stability and performance using the \ltwo-gain concept is not the proper method to also analyze performance requirements for NL systems through the LPV framework in case of reference tracking and disturbance rejection.

\section{Conclusion}\label{sec:Conclusion}
In this paper we showed that the notion of $\ltwo$-gain stability is not able to guarantee reference tracking and disturbance rejection requirements for NL systems through the LPV framework. Using $\ltwo$-gain based LPV synthesis methods can result in oscillatory behavior when performing reference tracking and disturbance rejection. A new synthesis procedure is proposed in the form of the \litwo LPV controller which is able to guarantee incremental gain stability and performance, ensuring that tracking and disturbance requirements \textit{are} attained. Which is also verified based on the simulation study of the position control problem of a Duffing oscillator. 
Moreover, compared to the methods proposed in literature, the proposed method allows for easy performance shaping through the LPV framework; has the additional benefit that it can be used for larger class of NL control problems and systems; and existing LPV synthesis methods can be used during the synthesis process, resulting in a controller with a larger degree of freedom. 

\bibliography{References} 

\begin{thebibliography}{19}
\providecommand{\natexlab}[1]{#1}
\providecommand{\url}[1]{\texttt{#1}}
\providecommand{\urlprefix}{URL }
\expandafter\ifx\csname urlstyle\endcsname\relax
  \providecommand{\doi}[1]{doi:\discretionary{}{}{}#1}\else
  \providecommand{\doi}{doi:\discretionary{}{}{}\begingroup
  \urlstyle{rm}\Url}\fi

\bibitem[{Apkarian and Adams(1998)}]{Apkarian1998}
Apkarian, P. and Adams, R.J. (1998).
\newblock {Advanced gain-scheduling techniques for uncertain systems}.
\newblock \emph{IEEE Transactions on Control Systems Technology}.

\bibitem[{Apkarian et~al.(1995)Apkarian, Gahinet, and Becker}]{Apkarian1995}
Apkarian, P., Gahinet, P., and Becker, G. (1995).
\newblock {Self-scheduled $H_\infty$ control of linear parameter-varying
  systems: a design example}.
\newblock \emph{Automatica}.

\bibitem[{Fromion et~al.(2001)Fromion, Monaco, and Normand-Cyrot}]{Fromion2001}
Fromion, V., Monaco, S., and Normand-Cyrot, D. (2001).
\newblock {The weighted incremental norm approach: From linear to nonlinear
  $H_\infty$-control}.
\newblock \emph{Automatica}.

\bibitem[{Fromion et~al.(1999)Fromion, Scorletti, and Ferreres}]{Fromion1999a}
Fromion, V., Scorletti, G., and Ferreres, G. (1999).
\newblock {Nonlinear performance of a PI controlled missile: An explanation}.
\newblock \emph{International Journal of Robust and Nonlinear Control}.

\bibitem[{Fromion and Scorletti(2003)}]{Fromion2003}
Fromion, V. and Scorletti, G. (2003).
\newblock A theoretical framework for gain scheduling.
\newblock \emph{International Journal of Robust and Nonlinear Control}.

\bibitem[{Koelewijn et~al.(2019)Koelewijn, T\'oth, Mazzoccante, and
  Nijmeijer}]{Koelewijn2019NLTR}
Koelewijn, P.J.W., T\'oth, R., Mazzoccante, G.S., and Nijmeijer, H. (2019).
\newblock Nonlinear tracking and rejection using linear parameter-varying
  control.
\newblock \emph{Submitted to IEEE Transaction on Control Systems Technology}.

\bibitem[{Koelewijn and Tóth(2019)}]{Koelewijn2019IncrGain}
Koelewijn, P.J.W. and Tóth, R. (2019).
\newblock \emph{Incremental gain of LTI systems}.
\newblock Technical Report TUE CS. Eindhoven University of Technology.

\bibitem[{Kwiatkowski et~al.(2006)Kwiatkowski, Boll, and
  Werner}]{Kwiatkowski2006AutomatedGA}
Kwiatkowski, A., Boll, M.T., and Werner, H. (2006).
\newblock {Automated Generation and Assessment of Affine LPV Models}.
\newblock \emph{Proc. of the 45th IEEE Conference on Decision and Control}.

\bibitem[{Leith and Leithead(1999)}]{Leith1999}
Leith, D.J. and Leithead, W.E. (1999).
\newblock {Input-output linearization by velocity-based gain-scheduling}.
\newblock \emph{International Journal of Control}.

\bibitem[{Lohmiller and Slotine(1998)}]{Lohmiller-And1998}
Lohmiller, W. and Slotine, J.J.J.E. (1998).
\newblock {On Contraction Analysis for Nonlinear Systems}.
\newblock \emph{Automatica}.

\bibitem[{Lohmiller and Slotine(2000)}]{Lohmiller2000}
Lohmiller, W. and Slotine, J.J.J.E. (2000).
\newblock {Control system design for mechanical systems using contraction
  theory}.
\newblock \emph{IEEE Transactions on Automatic Control}.

\bibitem[{Packard(1994)}]{Packard1994}
Packard, A. (1994).
\newblock {Gain scheduling via linear fractional transformations}.
\newblock \emph{Systems and Control Letters}.

\bibitem[{Pavlov et~al.(2004)Pavlov, Pogromsky, van~de Wouw, and
  Nijmeijer}]{Pavlov2004}
Pavlov, A., Pogromsky, A., van~de Wouw, N., and Nijmeijer, H. (2004).
\newblock {Convergent dynamics, a tribute to Boris Pavlovich Demidovich}.
\newblock \emph{Systems and Control Letters}.

\bibitem[{Pavlov et~al.(2007)Pavlov, van~de Wouw, and Nijmeijer}]{Pavlov2007}
Pavlov, A., van~de Wouw, N., and Nijmeijer, H. (2007).
\newblock {Global nonlinear output regulation: Convergence-based controller
  design}.
\newblock \emph{Automatica}.

\bibitem[{Scherer(2001)}]{Scherer2001}
Scherer, C.W. (2001).
\newblock {LPV control and full block multipliers}.
\newblock \emph{Automatica}.

\bibitem[{Scorletti et~al.(2015)Scorletti, Fromion, and {De
  Hillerin}}]{Scorletti2015}
Scorletti, G., Fromion, V., and {De Hillerin}, S. (2015).
\newblock {Toward nonlinear tracking and rejection using LPV control}.
\newblock \emph{Proc. of the 1st IFAC Workshop on Linear Parameter Varying
  Systems}.

\bibitem[{T{\'o}th(2010)}]{toth2010modeling}
T{\'o}th, R. (2010).
\newblock \emph{{Modeling and Identification of Linear Parameter-Varying
  Systems}}.
\newblock Springer.

\bibitem[{Wu(1995)}]{Wu1995}
Wu, F. (1995).
\newblock {Control of linear parameter varying systems}.
\newblock \emph{Ph.D. dissertation, University of Berkeley, California and
  USA}.

\bibitem[{Zames(1966)}]{Zames1966}
Zames, G. (1966).
\newblock {On the Input-Output Stability of Time-Varying Nonlinear Feedback
  Systems Part I: Conditions Derived Using Concepts of Loop Gain, Conicity, and
  Positivity}.
\newblock \emph{IEEE Transactions on Automatic Control}.

\end{thebibliography}

\end{document}